\documentclass[twocolumn,amsmath,amssymb,nofootinbib,tighten,floatfix,prl]{revtex4}

\usepackage{color}
\usepackage{graphicx}
\usepackage{dcolumn} 
\usepackage{bm, bbm}      
\usepackage{curves}
\usepackage{epic}
\usepackage{wasysym}
\usepackage{epsf, times}
\usepackage{subfigure}
\usepackage{eufrak}
\usepackage{amsthm}

\begin{document}

\title{Corner Junction as a Probe of Helical Edge States}

\author{Chang-Yu Hou$^{1}$}
\author{Eun-Ah Kim$^{2,3}$}
\author{Claudio Chamon$^{1}$}

\affiliation{$^{1}$\!\!\!
Physics Department, Boston University, Boston, MA 02215, USA
\\
$^{2}$\!\!\!
Department of Physics, Stanford University, Stanford, CA 94305, USA
\\
$^{3}$\!\!\!
Department of Physics, Cornell University, Ithaca, NY 14853, USA
}

\date{\today}

\begin{abstract}
We propose and analyze inter-edge tunneling in a quantum spin Hall corner junction as a means to probe the helical nature of the edge states. We show that electron-electron interactions in the one-dimensional helical edge states result in Luttinger parameters for spin and charge that are intertwined, and thus rather different than those for a quantum wire with spin rotation invariance. Consequently, we find that the four-terminal conductance in a corner junction has a distinctive form that could be used as evidence for the helical nature of the edge states.
\end{abstract}

\maketitle

\def\openone{\leavevmode\hbox{\small1\kern-4.2pt\normalsize1}}

\newcommand{\slapar}{\not \hskip -2 true pt \partial\hskip 2 true pt}
\newcommand{\slapartxt}{\not\!\!\!\!\partial}
\newcommand{\slaA}{\!\not\!\! A}
\newcommand{\slaAA}{\!\not\!\! A^{5}}
\newcommand{\slaM}{\ \backslash \hskip -8 true pt M}
\newcommand{\slaS}{\ \backslash \hskip -7 true pt \Sigma}
\newcommand{\beq}{\begin{equation}}
\newcommand{\eeq}{\end{equation}}
\newcommand{\bea}{\begin{eqnarray}}
\newcommand{\eea}{\end{eqnarray}}
%
%

\textit{Introduction} -- The better understanding of topological
phases in condensed matter physics, attained through the comprehensive
study of the quantum Hall (QH) effect, has led to the search for other
forms of topological states in the absence of applied magnetic
fields~\cite{Haldane88}. In particular, the quantum spin Hall (QSH) effect has been
proposed theoretically in various systems with time reversal (TR)
symmetry and spin-orbit
interactions~\cite{Onoda05,Kane05,Bernevig06,Qi06,Bernevig06-1}. A
recent experiment~\cite{Konig07} has provided evidence for transport
properties that are consistent with those associated with the QSH
effect: independence of the conductance from sample width, in line
with transport taking place at the edges, and sensitivity to an
external magnetic field, which breaks TR symmetry and destroys the
QSH.

The presence of a bulk gap and gapless edge states is a distinctive
signature of QSH insulators as new topological states of
matter~\cite{Haldane88,Kane05-Z2,Kane05,Roy06,Bernevig06,Fu,
Moore07}. For a two-dimensional (2D) system, these edge states are
expected to form a new type of 1D fermionic system, the ``helical
Luttinger liquid'' (HLL), where opposite spin modes counter
propagate~\cite{Wu06,Cenke06}. However, to the best of our knowledge,
experimental results or proposals for experiments that can directly
confirm the helical nature of the edge states and distinguish them
from ordinary Luttinger Liquids (LLs) are still lacking.

In this paper, we propose and analyze a {\it corner junction} with a
single point contact as a minimalistic but concrete setting for
probing the helical nature of the QSH edge states. In particular, we
find that the helicity constraint allows for a stable fixed point [in
the renormalization group (RG) sense] corresponding to a charge
insulator and spin conductor for tunneling across the point of
contact. This fixed point arises in the regime of sufficiently large
repulsive electron-electron interactions, and could be experimentally
accessed by choosing device parameters, such as the thickness of the
HgTe/(Hg,Cd)Te QSH insulator samples of
Refs.~\cite{Konig,Daumer,Becker07}. We derive the associated four
terminal linear conductance tensors through a formal ``folding
procedure'' which maps the corner junction of a pair of HLLs into a
junction of semi-infinite spinful Luttinger liquids (LLs).  The
non-trivial spin tunneling fixed point uniquely allows for a peculiar
flow of charges: currents flow into two terminals biased at
intermediate voltages not only from the terminal with the highest bias
but also {\it from the terminal with the lowest bias}!  Such four
terminal conductance, we argue, is a characteristic of a corner
junction of HLLs and can thus be used as unambiguous evidence for HLL
behavior of the QSH edge states.


\begin{figure}
\includegraphics[angle=0,scale=0.3]{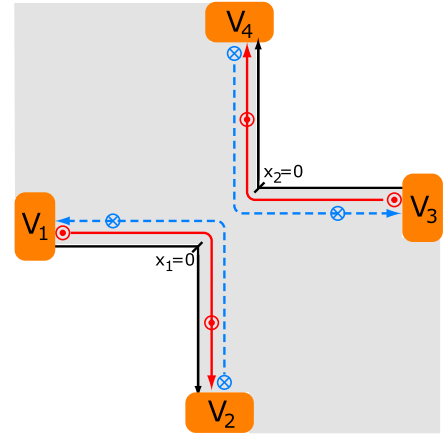}
\caption{(Color Online) Proposed geometry of a point contact device
for probing the helical nature of QSH edge states. The painted (grey)
area defines the bulk of the QSH insulator. The red (solid) line
represents the up-spin right-movers while the blue (dashed) line
represents the down-spin left-movers for both edges. Electrons can
tunnel between the two edges at $x_{1,2}=0$, where the corners
come into close proximity. There are four contact leads, where
voltages $V_{i}$ can be applied, before and after the tunneling point.
}
\label{fig:QSH-geometry}
\end{figure}


\textit{Geometry} -- The corner junction we propose, which can be
fabricated in a HgTe/(Hg,Cd)Te quantum well~\cite{Konig,
Daumer,Becker07}, is the four terminal geometry shown in
Fig.~\ref{fig:QSH-geometry}. The HLL edges are at the boundaries of
the bulk QSH insulator, which is shown in grey in the figure. The two
sets of HLL edge states (one running from terminal $1$ to $2$, and the
other from $3$ to $4$) are brought close to one another at their
corners, $x_{1,2}=0$, forming a point contact between two QSH edges,
where tunneling can occur. Four leads, where voltages $V_i$ can be
applied and currents $I_i$ can be measured ($i=1,2,3,4$), contact the
sample before and after the junction. The currents are defined as
positive when flowing out of the leads and into the edges. Our
discussion will focus only on the effects attributed to the point
contact. However, in a QSH sample, there should be additional pairs of
HLL edges, one that connects leads 1 and 4, and another that connects
leads 2 and 3. For extracting the significant features due to the
point contact alone, one can either isolate the region of interest
with additional contacts or take the contribution of the extra HLLs
into account by including a conductance $e^2/h$ between the
appropriate leads.


\textit{Folded picture of the helical Luttinger liquids} -- The HLL
has only half the degrees of freedom of a conventional 1D
system~\cite{Wu06,Cenke06}, because helicity (preserved under TR
symmetry) correlates spin polarization with the direction of
propagation. This helicity distinguishes the HLL from other states
where the degrees of freedom are reduced by half, such as the spinless
LL and the chiral LL, whose chirality is induced by a magnetic field
necessary for the QH effect. Consider a HLL consisting of right (left)
movers $\psi_{R \uparrow}$ ($\psi_{L\downarrow}$) that carry up
(down)-spin. The linearized Hamiltonian of the HLL, in its
non-interacting limit, can be cast as
\begin{equation}
\label{eq:non-inter-Hamil}
H_0=- v_F \int dx \left( \psi^\dag_{R \uparrow} 
\,i \partial_x \psi^{\,}_{R \uparrow} 
- \psi^\dag_{L \downarrow} 
\,i \partial_x \psi^{\,}_{L
\downarrow} \right),
\end{equation}
where 
TR symmetry forbids all TR odd perturbations: single particle
backscattering operators, which open up a mass gap, are thus
excluded~\cite{Wu06,Cenke06}. The chiral interaction for the same species can
be written as
\begin{equation}
\label{eq:chiral-interaction-Hamil}
H_{\rm ch}
=\frac{\lambda_4}{2} \int dx \left( 
\psi^\dag_{R\uparrow} \psi^{\,}_{R\uparrow} 
\psi^\dag_{R\uparrow} \psi^{\,}_{R\uparrow} 
+\psi^\dag_{L\downarrow} \psi^{\,}_{L\downarrow} 
\psi^\dag_{L\downarrow} \psi^{\,}_{L\downarrow} 
\right),
\end{equation}
where $\lambda_4$ is the interaction constant. There are two TR
invariant non-chiral interactions, the forward scattering and the
Umklapp scattering. We shall neglect the Umklapp scattering, which is
important only for certain commensurate fillings~\cite{Wu06,
Cenke06}. The Hamiltonian for the forward scattering reads
\begin{equation}
\label{eq:forward-scattering-Hamil}
H_{\rm fw} = \lambda_2 \int dx\left( \psi^\dag_{R\uparrow}
\psi^{\,}_{R\uparrow} \psi^\dag_{L\downarrow} \psi^{\,}_{L\downarrow}
\right),
\end{equation}
with $\lambda_2$ as the interacting constant. Observe that the spin
degrees of freedom are redundant, hence one can effectively treat the
HLL as a spinless LL system and define the boson fields
$\varphi=\phi_{R\uparrow}+\phi_{L\downarrow}$ and $\theta=
\phi_{L\downarrow} - \phi_{R\uparrow}$ within the standard
bosonization procedure. The bosonized Hamiltonian, $H=H_0+H_{\rm
ch}+H_{\rm fw}$, reads
\begin{equation}
\label{eq:Hamiltonian}
H=\frac{v}{4 \pi }\int dx \left[  \frac{1}{g} (\partial_x \theta)^2 
+ g (\partial_x \varphi)^2 \right]
\;,
\end{equation}
where the velocity $v\equiv v_F\sqrt{(1+\frac{\lambda_4}{2 \pi v_F}
)^2-( \frac{\lambda_2}{2\pi v_F})^2 }$ and the Luttinger parameter
$g\equiv \sqrt{\frac{2\pi v_F+\lambda_4-\lambda_2}{2\pi v_F+\lambda_4
+\lambda_2}}$. Hence, the behavior of a HLL consisting of one pair of
edge states is controlled by a Luttinger parameter $g$ and it is
similar to a spinless LL. However, unlike a spinless LL, a HLL is
protected from localization by TR symmetry, which forbids
backscattering.

Although the HLL is effectively a spinless LL in an infinite wire, for
the particular corner junction geometry depicted in
Fig.~\ref{fig:QSH-geometry}, it is convenient to map the HLL into a
spinful LL in a semi-infinite wire. Let us introduce the following
mapping for the edge states indexed by $\alpha=1,2$ that come to the
corner at $x_{\alpha}=0$
\begin{equation}
\psi^{\alpha}_{L\downarrow}(x_\alpha) 
\to \psi^{\alpha}_{R \downarrow}(-x_ \alpha),\;
\psi^{\alpha}_{R\uparrow}(x_ \alpha) 
\to \psi^{\alpha}_{L \uparrow}(-x_ \alpha),\;
\forall\; x_\alpha < 0.
\end{equation}
All fields ($\psi^{\alpha}_{R\uparrow}, \psi^{\alpha}_{R\downarrow},
\psi^{\alpha}_{L\uparrow}$ and $\psi^{\alpha}_{L\downarrow}$) are thus
effectively defined for a semi-infinite ($x_\alpha>0$) wire, owing to
a proper boundary condition (BC). Then, the charge and spin boson
fields in the standard bosonization scheme are defined as
\begin{equation}
\label{eq:charge-spin-fields}
\varphi_c=\frac{1}{\sqrt{2}} (\varphi_\uparrow + \varphi_\downarrow );
\qquad \varphi_s=\frac{1}{\sqrt{2}} (\varphi_\uparrow -
\varphi_\downarrow )
\end{equation}
and likewise for the dual fields $\theta_{c(s)}$, where
$\varphi_{\sigma}=\phi_{R,\sigma}+\phi_{L,\sigma}$ and $\theta_\sigma=
\phi_{L,\sigma}- \phi_{R,\sigma}$ for $\sigma=\uparrow,
\downarrow$. Finally, the Hamiltonian for the two copies
($\alpha=1,2$) of edge states can be written as
\begin{equation}
\label{eq:folded-Hamil}
H=\sum_{a, \alpha} \frac{v_a}{4 \pi } \int_{x>0} dx \left[ \frac{1}{g_a}
(\partial_x \theta^ \alpha_a)^2 + g_a (\partial_x \varphi^ \alpha_a)^2 \right],
\end{equation}
where $a=c,s$ represent the charge and spin degrees of freedom, the
Luttinger parameters $g_c=g=(g_s)^{-1}$, and the normalized velocity
$v_c=v_s=v$. Hence, the folding procedure we describe above maps the
corner junction between a pair of HLLs into a junction of two
semi-infinite spinful LLs with the constraint, $g_c\!\times\!
g_s\!=\!1$, a manifestation of the helical nature of the QSH edge
states. This relation is in stark contrast to the one in the simple
LL, for which $g_s\!=\!1$ is required by spin rotation
invariance.



\begin{figure}
\includegraphics[angle=0,scale=0.55]{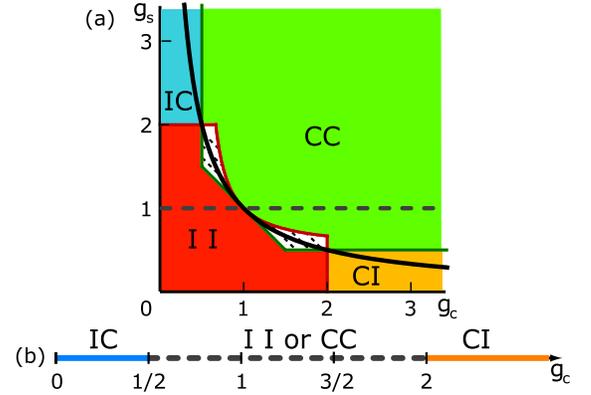}
\caption{(Color online) (a) The phase diagram of a junction of two
spinful LL quantum wires. The black curve indicates the trajectory
that respects the HLL constraint $g_c\!\times\! g_s\!=\!1$. The label
C stands for conducting while the label I stands for insulating
phase. The adjacent labeling, AB, indicates the phase A and phase B
for the charge and spin degrees of freedom respectively. Depending on
the detailed structure of the point contact, both CC and II fixed
points can be stable at low energies, and this is represented in the
dashed area. (b) shows the phase diagram of the point contact between
two HLL edge states in terms of $g_c$, obtained by following the
$g_c\!\times\! g_s\!=\!1$ trajectory in (a).
}
\label{fig:N-2-phase}
\end{figure}

\textit{Low energy fixed points} -- The folding procedure allows us to
easily identify the fixed points for the corner junction (see
Fig.~\ref{fig:QSH-geometry}). For arbitrary values of $g_c, g_s$, a
spinful LL with a single tunneling center was analyzed using
perturbative RG by Kane and Fisher~\cite{Kane92} and in terms of
boundary conditions imposed by the tunneling center by Wong and
Affleck~\cite{Wong}. One can take the $g_c\!\times\! g_s\!=\!1$
parametric line and follow it on the $(g_c,g_s)$ plane, and use the
results for the spinful LL in Ref.~\cite{Kane92, Wong}. As depicted in
Fig.~\ref{fig:N-2-phase}(a), transport through the point contact is
renormalized to the charge conductor and spin insulator (CI) fixed
point when $g_c >2$, while it is renormalized to the charge insulator
and spin conductor (IC) fixed point when $g_c <1/2$. In between these
phases, when $1/2<g_c<2$, the system can be either a charge and spin
conductor (CC) or a charge and spin insulator (II), depending on the
detailed structure of the point contact. The phase diagram of the
system as a function of $g_c$ is plotted in
Fig.~\ref{fig:N-2-phase}(b).

Strikingly, two fixed points, IC and CI, which are unattainable for an
ordinary LL with $g_s=1$, are accessible for the HLL by tuning the
interaction parameter $g_c$. In particular, transport through the
corner junction of two HLL renormalizes into the IC low energy fixed
point with strong repulsive interaction, $g<1/2$, that can be
engineered, as estimated below, with proper experimental parameters in
the HgTe/(Hg,Cd)Te quantum well. Consequently, by measuring the charge
transport properties of a corner junction of two HLL edge states, one
can clearly discriminate between the HLL and the LL: in the latter,
renormalization {\it always} results in the totally disconnected fixed
point, {\it i.e.}  {\it no current flow} across the tunneling junction
when interactions are repulsive, $g<1$. Also, it is worthwhile to
compare the corner junction to a similar experimental setup, with a
single point contact between two chiral LL edge states supported by
the QH liquid. There, depending on the details of the point contact
and the filling fraction, only two phases, corresponding to the weak
and the strong tunneling limit, are possible. Hence, the
renormalization into the IC fixed point for strong repulsive
interactions is a distinctive hallmark of the HLL.

It is rather non-trivial to calculate the actual value of $g$ from
microscopic parameters. However, the relation $g\approx
\frac{1}{\sqrt{1+U/(2E_F)}}$, where $E_F$ is the Fermi energy and $U$
is the characteristic Coulomb energy of the system~\cite{Kane92},
provides a rough estimate. For a HgTe/CdTe quantum well with width
$w=7$~nm, $U\approx e^2/w \approx 0.2$ eV, and $E_F$ is approximately
the band gap $40$~meV~\cite{Becker07}. Then, the Luttinger parameter
is estimated to be $g \approx 0.53$ and can be adjusted by changing
the width of the quantum well.



\textit{Conductance tensor} -- Now we discuss how the terminal
transport measurements can reveal the nature of the different fixed
points. The conductance tensor, $G_{ij}$, defined as the current
response to the applied voltage $I_{i}=G_{ij} V_j$, can be calculated
using the Kubo formula with the proper identification of the
conformally invariant BCs associated to the low energy fixed
points~\cite{Oshikawa, Hou}.

Generically, the BCs can be encoded into rotation matrices
$\mathcal{R}_{c(s)}$ that relate the left and right movers, of the
charge and spin degrees of freedom separately, through
$\phi^{\alpha}_{R,c(s)} = \mathcal{R}^{\alpha \beta}_{c(s)}
\phi_{L,c(s)}^\beta$, where $\alpha,\beta=1,2$ are the edge
indices. Notice that only combinations of two types of BCs appear in
this problem: insulator (Neumann) BC and conductor (Dirichlet) BC, and
the corresponding rotation matrices are given by
\begin{equation}
\label{eq:BC-rotation-matrix}
\mathcal{R}^{I}_{c(s)}= 
\left(\begin{array}{cc}1 & 0 
\\
0 & 1\end{array}\right); 
\qquad \mathcal{R}^{C}_{c(s)}= \left(\begin{array}{cc}0 & 1 
\\
1 & 0\end{array}\right).
\end{equation}
However, for computing $G_{ij}$ for each BC, it is
most convenient to identify  the $4 \times 4$ rotation
matrix that relates the incoming fields
$\Phi_I=(\phi^1_{L,\uparrow},\phi^1_{L,\downarrow},
\phi^2_{L,\uparrow},\phi^2_{L,\downarrow})^T$ to the outgoing fields
$\Phi_O=(\phi^1_{R,\downarrow},\phi^1_{R,\uparrow},
\phi^2_{R,\downarrow},\phi^2_{R,\uparrow})^T$ in each channel, for a given BC: 
$\Phi_O=\mathcal{R}^{BC} \Phi_{I}$. Then, the conductance tensor
$G_{ij}$ can be derived from the Kubo formula and written in 
a compact form~\cite{Hou, Bellazzini}
\begin{equation}
\label{eq:Conductance}
G^{BC}_{ij}=g \frac{e^2}{h}(\delta_{ij}-\mathcal{R}^{BC}_{ij})
\;.
\end{equation}

After some algebra, the rotation matrices $\mathcal{R}^{IC}$ and
$\mathcal{R}^{CI}$ for IC and CI BCs can be derived from the
combinations of $\mathcal{R}^{I,C}_{c(s)}$ in Eq.~(\ref{eq:BC-rotation-matrix}) as
\begin{equation}
\left(\!\!\!
\begin{array}{cccc}
 \;\;\;\frac{1}{2} & \;\;\;\frac{1}{2} & -\frac{1}{2} & \;\;\;\frac{1}{2} \\
 \;\;\;\frac{1}{2} & \;\;\;\frac{1}{2} & \;\;\;\frac{1}{2} & -\frac{1}{2} \\
 -\frac{1}{2} & \;\;\;\frac{1}{2} & \;\;\;\frac{1}{2} & \;\;\;\frac{1}{2} \\
 \;\;\;\frac{1}{2} & -\frac{1}{2} & \;\;\;\frac{1}{2} & \;\;\;\frac{1}{2}
\end{array}
\right)
\;
{\rm and}
\;
\left(\!\!\!
\begin{array}{cccc}
 -\frac{1}{2} & \;\;\;\frac{1}{2} & \;\;\;\frac{1}{2} & \;\;\;\frac{1}{2} \\
 \;\;\;\frac{1}{2} & -\frac{1}{2} & \;\;\;\frac{1}{2} & \;\;\;\frac{1}{2} \\
 \;\;\;\frac{1}{2} & \;\;\;\frac{1}{2} & -\frac{1}{2} & \;\;\;\frac{1}{2} \\
 \;\;\;\frac{1}{2} & \;\;\;\frac{1}{2} & \;\;\;\frac{1}{2} & -\frac{1}{2}
\end{array}
\right),
\end{equation}
respectively. Using Eq.~(\ref{eq:Conductance}), the conductance tensors at zero temperature and zero bias for the IC and CI fixed points are given by
\begin{subequations}
\label{eq:conductance-tensor}
\begin{eqnarray}
\label{eq:conductance-tensor-IC}
G^{IC}&=& \frac{g e^2}{2 h} \left(\begin{array}{cccc}
\;\;\;1 & -1 & \;\;\;1 & -1 \\
-1 & \;\;\;1 & -1 & \;\;\;1 \\
\;\;\;1 & -1 & \;\;\;1 & -1 \\
-1 & \;\;\;1 & -1 & \;\;\;1
\end{array}\right),
\\
\label{eq:conductance-tensor-CI}
G^{CI}&=&\frac{g e^2}{2 h} \left(\begin{array}{cccc}
\;\;\;3 & -1 & -1 & -1 \\
-1 & \;\;\;3 & -1 & -1 \\
-1 & -1 & \;\;\;3 & -1 \\
-1 & -1 & -1 & \;\;\;3
\end{array}\right).
\end{eqnarray}
\end{subequations}
Observe that these conductance tensors satisfy the constraints
$\sum_{i} G_{i j }=0$ and $\sum_j G_{i j} =0$ due to current
conservation and to the fact that currents vanish when all four
applied voltages are equal, respectively. Also, with the contact
resistance between the leads and the edge states taken into account,
the overall conductance will take the same form as in
Eq.~(\ref{eq:conductance-tensor}) but with a simple substitution
$g\to 1$. Notice that the conductance tensors of the IC
and CI fixed points are not block diagonal and show the twined
response of the four terminals, the exclusive features that can be
used to detect the helical nature of the edge states.

\begin{figure}
\includegraphics[angle=0,scale=0.35]{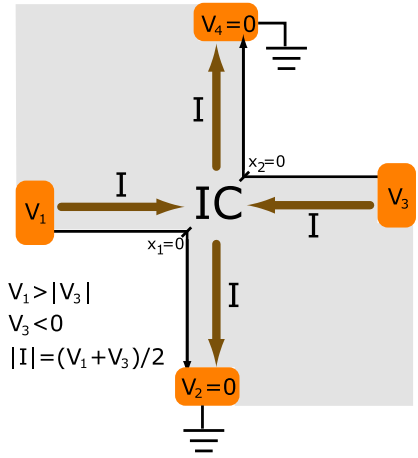}
\caption{(Color online) Proposed measurement for detecting the IC
fixed point and the helical nature of the QSH edge states. A positive
voltage $V_1$ and a negative voltage $V_3$ are applied to leads 1 and
3, such that $V_1>|V_3|$, while leads 2 and 4 are grounded. The arrows
indicate the direction of the current flow and the quantized
conductance $e^2/h=1$. Notice that currents flow {\it out} from lead
3, even though it has the {\it lowest} applied voltage.}
\label{fig:setup}
\end{figure}

\textit{Proposed measurement} -- Perhaps the most remarkable
consequence of the IC fixed point is the possibility that a current
can flow out from the lead with the lowest applied voltage. [Notice
that this does not violate thermodynamic principles, since the
dissipated power at the junction can only be non-negative, as the
eigenvalues of the conductance tensors in
Eq.~(\ref{eq:conductance-tensor}) are larger or equal to zero.]
Specifically, in the setup shown in the Fig.~\ref{fig:setup}, when a
positive voltage $V_1$ and a negative voltage $V_3> -V_1$ are applied
to terminals 1 and 3, respectively, with leads 2 and 4 grounded, a
current $I_3=e^2(V_3+V_1)/2h$ will flow {\it out} from lead 3, which
has the lowest applied voltage! This counterintuitive result would be
a smoking gun evidence, out of the corner junction measurement, of the
helical nature of the edge states supported by the QSH insulator.

As in the case of other types of LLs, the currents through the corner
junction of HLLs will acquire power law corrections at finite
temperatures and voltages: $\delta G(T)\sim T^{2(\Delta_{\rm min}-1)}$
and $\delta I(V)\sim V^{2\Delta_{\rm min}-1}$~\cite{Kane92}.
$\Delta_{\rm min}$ is the scaling dimension of the leading irrelevant
boundary operators that tend to drive the system away from the fixed
point, and varies according to the fixed point and Luttinger
parameter. At the IC fixed point, we find that $\Delta_{\rm
min}=1/(2g_c)$ for all range of $g_c<1/2$. At the II and CC fixed
points, $\Delta_{\rm min}=2 g_c$ or $\Delta_{\rm
min}=(g_c+g_c^{-1})/2$, depending on the value of $g_c$ within the
interval $1/2<g_c<1$. Since the II and CC fixed points are most likely
realized in weakly interacting limit, examining the power law
corrections in transport data is already a step towards demonstration
of HLLs through corner junctions, which can be more explicitly shown
in the strongly interacting regime with the IC fixed point.


\textit{Broken $S_z$ symmetry} -- Throughout our discussions, we have
assumed that the polarization of the spins in the two edges are the
same, which can be achieved in a HgTe quantum well sample if the spin
polarization is tied to the crystaline directions. However, the
polarization of spin in two edges can be in general different, and the
stability of the low energy fixed point may be altered accordingly. We
found that the II, CC and CI fixed points still remain stable. On the
other hand, tunneling processes that were originally forbidden due to
the spin conservation destabilize the IC fixed point. Moreover, we are
unable to identify the stable fixed point when $g_c<1/2$. For this
range of $g_c$, we now find three unstable fixed points (including the
IC fixed point), each unstable along the direction pointing to the
other two, which suggest the existence of intermediate fixed points
which are not obviously accessible using the methods we used here. The
origin of the stable fixed point when $S_z$ is not a good quantum
number is an interesting open problem.


\textit{Summary} -- We proposed and analyzed a corner junction in a
QSH insulator as a simple yet rather effective test bed of the helical
properties of the edge states and the non-trivial topological nature
of the QSH insulator. By mapping the corner junction of HLLs connected
to four reservoirs into a junction of two spinful LL quantum wires, we
found that an unmistakable IC fixed point is accessible when
electron-electron interactions are sufficiently repulsive. This fixed
point can be attained by engineering HgTe/HgCd quantum wells so as to
enhance the repulsive interactions within a single HLL. The four
terminal conductance tensor associated to the IC regime has a telltale
sign: currents can flow {\it out} of a reservoir with the {\it lowest}
bias (this is possible, without violating thermodynamic principles,
because currents flows among four and not only two terminals). If
experimentally observed, this unique conductance tensor can provide
unambiguous evidence for the helical nature of the edge states at the
boundaries of topological QSH insulators.

We thank L.~Molenkamp and S.~Ryu for enlightening discussions. This
work is supported in part by the DOE Grant
DE-FG02-06ER46316~(C-Y.~H. and C.~C.) and by the Stanford Institute
for Theoretical Physics~(E-A.~K.).

\end{document}